\documentclass[fleqn,twoside]{article}
\usepackage{espcrc2}
\usepackage{amssymb}

\usepackage{graphicx}

\newcommand{\be}{\begin{equation}}
\newcommand{\ee}{\end{equation}}
\newcommand{\bea}{\begin{eqnarray}}
\newcommand{\eea}{\end{eqnarray}}

\newcommand{\chiPT}{$\chi$PT}

\newcommand{\mpi}{m_\pi}

\newcommand{\gev}{\, {\rm GeV}}
\newcommand{\mev}{\, {\rm MeV}}
\newcommand{\fm}{\, {\rm fm}}

\title{
\vspace{-2.6cm}
\hfill \rm \null \hfill
\hbox{\normalsize ADP-03-140/T575} \\
\vspace{1.85cm}
Chiral extrapolation and physical insights}

\author{	R.~D.~Young,
		D.~B.~Leinweber and
		A.~W.~Thomas
		\\ \vspace*{3mm}
		Special Research Centre for the
		Subatomic Structure of Matter,
		and Department of Physics,
		University of Adelaide, Adelaide SA 5005,
		Australia.}

\begin{document}

\begin{abstract}
  It has recently been established that finite-range regularisation in
  chiral effective field theory enables the accurate extrapolation of
  modern lattice QCD results to the chiral regime. We review some of
  the highlights of extrapolations of quenched lattice QCD results,
  including spectroscopy and magnetic moments. The $\Delta$ resonance
  displays peculiar chiral features in the quenched theory which can
  be exploited to demonstrate the presence of significant chiral
  corrections.
\vspace{1pc}
\end{abstract}

\maketitle

\section{INTRODUCTION}
Over the past few years there has been a significant effort into the
study of the chiral extrapolation problem for lattice QCD.
Extrapolation of lattice QCD simulation results to the light quark
mass regime is nontrivial due to nonanalytic variation of hadron
properties with quark mass. Such nonanalytic behaviour arises as a
consequence of spontaneously broken chiral symmetry in the QCD ground
state.

Early research in extrapolations of nucleon properties
\cite{Leinweber:1998ej,Leinweber:2000ig,Detmold:2001jb} discovered
that one could obtain reliable extrapolations by incorporating the
model-independent constraints of chiral symmetry near the chiral
limit.  Importantly, these studies highlighted the fact that the
rapid, nonanalytic variation with quark mass must be suppressed at
some intermediate energy scale. Above pion masses $\sim 500\mev$,
nucleon properties become smoothly varying functions of the quark
mass, $m_q\propto \mpi^2$. This observation has a simple physical
explanation. Once the Compton wavelength of the pion becomes small
($\lambda_C\sim\mpi^{-1}$) relative to the finite extent of the
nucleon, chiral loop effects become suppressed \cite{Detmold:2001hq}.

Recent work has established that these features are all naturally
built into chiral perturbation theory (\chiPT) when evaluated with a
finite-range regulator (FRR) \cite{Young:2002ib}. The formulation of
\chiPT\ with chiral loop integrals cut-off in momentum space at a
finite energy scale has been established by Donoghue {\it et al.}
\cite{Donoghue:1998bs}. Finite-range cutoff effects in chiral
effective field theory have also been studied in the context of
lattice regularised \chiPT\ \cite{Borasoy:2002hz}.

A case study of the nucleon mass expansion \cite{Young:2002ib} has
demonstrated that the use of FRR improves the convergence properties
of the \chiPT\ expansion.  The expansion has been demonstrated to be
applicable up to pion masses of at least $\mpi^2\sim 0.8\gev^2$. This
allows a reliable connection with modern lattice QCD simulations. A
range of functional forms for the momentum space cutoff have been
investigated and the systematic errors induced by the cutoff have been
found to be below the 1\% level.

\begin{figure}[t]
\includegraphics[width=\columnwidth]{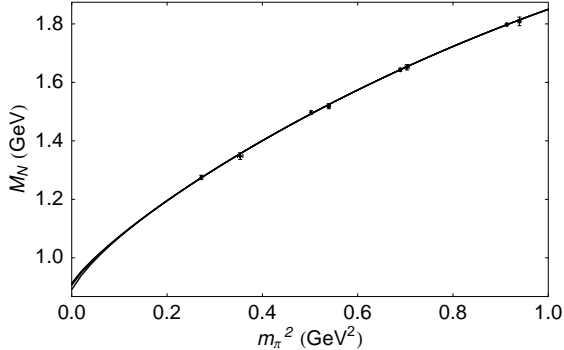}
\caption{Extrapolation of nucleon mass lattice data using four
  different finite-range regulators \cite{Leinweber:2003dg}. The
  physical nucleon mass is {\bf not} a constraint in these fits.}
\label{fig:mNext}
\end{figure}

Following this study it has been demonstrated that one-loop
FRR-\chiPT\ is able to extrapolate modern lattice QCD simulations of
the nucleon mass with minimal systematic uncertainty
\cite{Leinweber:2003dg}.  Figure~\ref{fig:mNext} shows extrapolations
of two-flavour dynamical QCD lattice data \cite{AliKhan:2001tx} based
upon four different functional forms for the regulator.
The residual dependence on the regulator is beyound the resolution of
this figure.
With an increasing number of dynamical baryon simulations in lattice
QCD \cite{AliKhan:2001tx,Bernard:2001av,Allton:2001sk,Aoki:2002uc} it
is essential to incorporate a correct treatment of chiral physics.
Only then can an interesting comparison of theory and experiment be
carried out.

In this paper we highlight some of the features learned about the
chiral structure of baryons. In particular we look at the nucleon and
the $\Delta$ resonance in both dynamical and quenched simulations of
QCD.  The majority of the effects of dynamical sea quarks in baryon
masses, relative to QQCD, can be explained by the modified chiral
structure \cite{Young:2002cj}. Special features of the $\Delta$ baryon
in quenched QCD also allow for a unique opportunity to study quenched
chiral physics \cite{Leinweber:2003ux,Young:2003gd}.

\section{BARYON MASSES}
As outlined in the Introduction the use of a FRR in \chiPT\ is
essential for the chiral expansion to be applicable at the quark
masses simulated in state of the art dynamical lattice simulations.
The application of quenched \chiPT\ \cite{Labrenz:1996jy} to quenched
simulations has demonstrated similar success with the use of a FRR
\cite{Young:2002cj}. The remarkable discovery of this work was that
the primary difference between quenched and dynamical spectroscopy can
be explained by the differences in the chiral loop contributions.

\begin{figure}[t]
\mbox{
$\Sigma_N = $
\hspace{1mm}
\includegraphics[width=0.8\columnwidth]{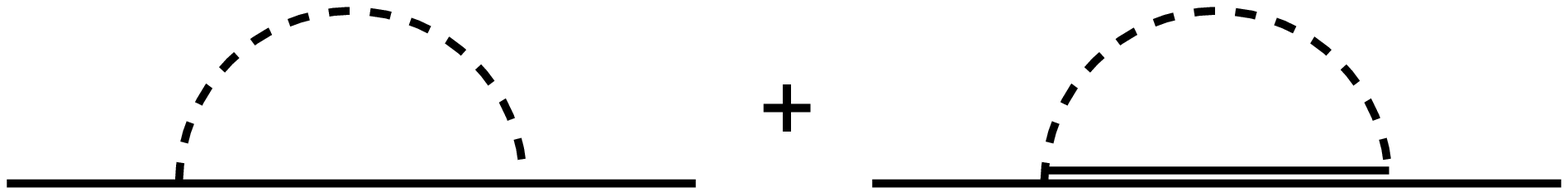}
}

\vspace*{6mm}

\mbox{
\put(0,35){$\Sigma_N^{\rm (Q)} = $}
\hspace{12mm}
\includegraphics[width=0.8\columnwidth]{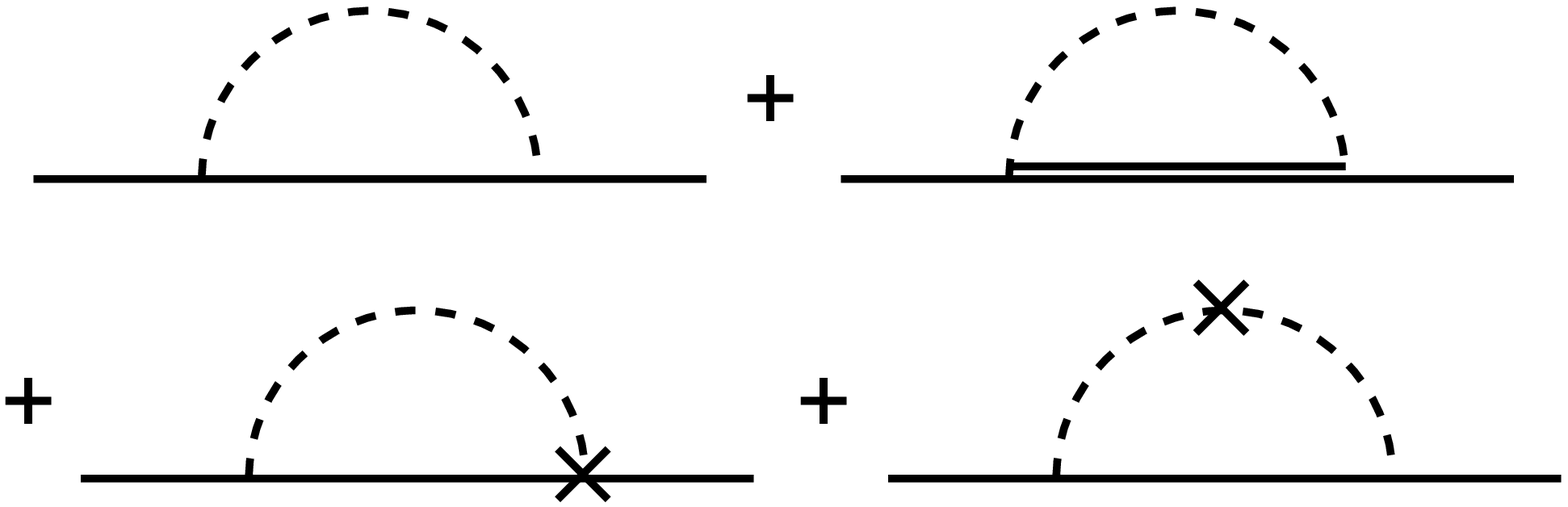}
}
\caption{Leading chiral loop corrections to the nucleon mass in QCD
  and quenched QCD. Solid, double and dashed curves correspond to
  nucleon, $\Delta$ and meson propagators. The cross denotes a hairpin
  insertion in the $\eta'$ propagator.}
\label{fig:se}
\end{figure}

Using FRR we express the chiral expansion of the nucleon mass as
\cite{Young:2002ib}
\begin{equation}
m_B = a_0 + a_2 \mpi^2 + a_4 \mpi^4 + \Sigma_B(\mpi,\Lambda) \, ,
\label{eq:mass}
\end{equation}
where $\Sigma_B$ is the total contribution to the nucleon mass from
chiral-meson loop diagrams and $\Lambda$ is a regulator parameter
governing the range. For the nucleon mass the corresponding
contributions are shown in Fig.~\ref{fig:se}, with analogous diagrams
for the $\Delta$. These diagrams give rise to the leading and
next-to-leading nonanalytic behaviour in the chiral expansion. It is
important to note that although $\Sigma_B$ is a function of $\Lambda$,
the dependence on $\Lambda$ is removed through renormalisation of the
chiral expansion \cite{Young:2002ib}.

In the quenched case we first note the inclusion of two additional
diagrams which are absent in the case of QCD. The flavour-singlet
$\eta'$ is also a light degree of freedom in the quenched theory and
hence contributes to the low-energy effective field theory
\cite{Labrenz:1996jy}. Details of the role of the $\eta'$ in the
context of baryon mass extrapolation have been discussed extensively
in Ref.~\cite{Young:2002cj}.

The contributions from the standard $\pi$-meson loops are also
modified in the quenched theory. In QCD meson-baryon loops are
attractive and hence the binding energy has the effect of lowering the
effective mass of the state in question. Of particular interest in the
quenched theory is the $N$--$\pi$ loop contribution to the $\Delta$
mass. The sign of this diagram is reversed and the loop acts
repulsively, raising the mass of the resonance in the quenched theory.

\begin{figure*}[tb]
\begin{center}
{\includegraphics[height=12cm,angle=90]{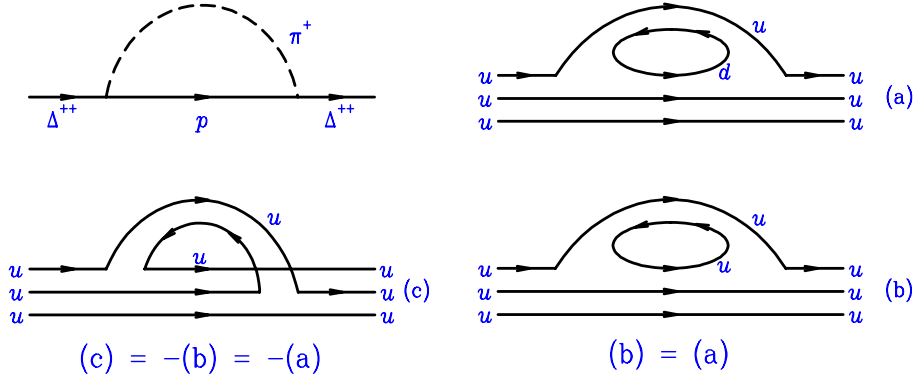}}
\caption{Pictorial quark-flow view of the $N$--$\pi$ meson-cloud 
contributions to the $\Delta$ baryon in full QCD.}
\label{deltaFlow}
\end{center}
\end{figure*}

This phenomenology can easily be understood by considering a
pictorial, quark-flow description of the processes contributing to the
$N$--$\pi$ loop in full QCD. Figure~\ref{deltaFlow} shows all
different topological contributions to the $\Delta^{++}\to N\pi$ loop
diagram. Diagrams (b) and (c) both indicate the propagation of an
unphysical $uuu$ octet baryon. These two processes must sum to zero in
the physical theory, {\it i.e.}~(c)=$-$(b). QCD is also flavour-blind and knows
nothing of the quark flavour propagating in the loop, meaning that
processes (a) and (b) are identical. In the quenched theory the only
remaining diagram is (c), which is a $uuu$ quark state degenerate with
the nucleon. By deduction this contribution is precisely the same in
magnitude as Fig.~\ref{deltaFlow}(a) with the opposite sign.

Fits of both quenched and dynamical simulations based upon
Eq.~(\ref{eq:mass}) have been performed in Ref.~\cite{Young:2002cj}.
In fitting to the results of staggered fermion simulations in quenched
and 2+1-flavour QCD \cite{Bernard:2001av} it was found that the
analytic coefficients $a_i$ of Eq.~(\ref{eq:mass}) show exceptional
agreement between the quenched and dynamical simulations. This
indicates that the most significant difference between the quenched
and dynamical spectroscopy can be accounted for by the corresponding
differences in the leading chiral corrections,
$\Sigma_B-\Sigma_B^{(Q)}$.

This observation allows one to take quenched lattice QCD data and
determine a phenomenological estimate of the quenching effects. One
simply fits quenched data using Eq.~(\ref{eq:mass}) with the
appropriate quenched chiral corrections. By retaining the fit
parameters $a_i$ and replacing the chiral corrections by their
corresponding full QCD contributions the physical masses can be
reconstructed.

\begin{figure}[tb]
\begin{center}
{\includegraphics[width=\columnwidth]{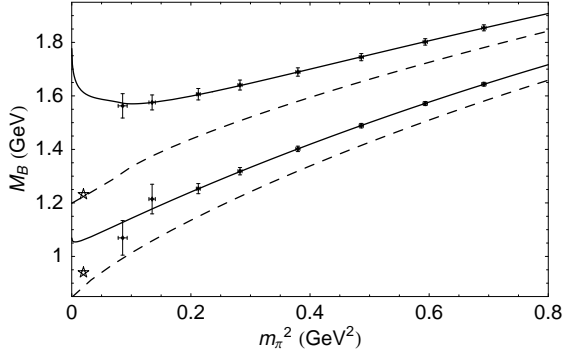}}
\caption{Solid curves show quenched fits to preliminary FLIC fermion lattice 
data. Dashed curves display estimates of the physical theory obtained
by unquenching the chiral loops. Stars indicate experimental values.}
\label{fig:flic}
\end{center}
\end{figure}

In Fig.~\ref{fig:flic} we show the fits to preliminary results of the
CSSM Lattice Collaboration using FLIC fermions \cite{Zanotti:2001yb}.
The lattice scale has been set via the Sommer scale, $r_0=0.5\fm$.
This gives a lattice spacing of $0.128\fm$ on a $20^3\times 40$
volume. The flattening of the $\Delta$ in the chiral regime is clearly
displayed by these new simulation results. This effect is primarily
caused by the reversal in sign of the $\Delta\to N\pi$ contribution as
discussed above.

The dashed curves in Fig.~\ref{fig:flic} represent our estimate of
unquenching effects where $\Sigma_B^{(Q)}$ has been replaced by
$\Sigma_B$. The chiral contributions here have been evaluated using a
$0.8\gev$ dipole regulator. This value was the optimal $\Lambda$
found in Ref.~\cite{Young:2002ib} which gave best agreement between
dynamical lattice QCD and experiment. These results are preliminary
and finite volume effects are yet to be taken into account, yet we
already see remarkable agreement with the experimentally measured
masses.

We see from Fig.~\ref{fig:flic} that the $\Delta$ suffers from larger
quenching effects due to the repulsive $N\pi$ loop contribution. It
has also been recognised that other excited states seem to lie
systematically high in the quenched approximation. Using chirally
improved quark actions the BGR Collaboration have noted that there
seems to be a correlation between the quenched discrepancies and the
physically observed widths \cite{Gattringer:2003qx}. As chiral
extrapolation techniques for excited states are developed
\cite{Morel:2003fj} it will be interesting to see if more quenched
spectroscopy can be similarly understood in terms of chiral
interactions.

\section{MAGNETIC MOMENTS}
Chiral loop effects are known to give large contributions to the
electromagnetic structure of baryons. In fact the leading nonanalytic
contribution to the neutron's magnetic moment is nearly one third the
experimental value.  Chiral symmetry effects in nucleon
electromagnetic structure describe dynamical, nonperturbative
structure beyond the simple quark model of QCD
\cite{Leinweber:2001ui}.  Such is the magnitude of the pion loop
effects that one should expect dramatic results in lattice
calculations as the light quark mass regime is probed.

Once again we find that the repulsive interaction of the $N\pi$ loop
means that the magnetic moment of the $\Delta$ baryon is especially
interesting in QQCD.

The best lattice calculations of the electromagnetic structure of
baryons have so far been restricted to the quenched model of QCD
\cite{Leinweber:1990dv,Gockeler:2003ay}.  Because of the absence of
$q\bar{q}$ pair-creation, pion loop effects are typically suppressed
in the quenched approximation
\cite{Savage:2001dy,Leinweber:2001jc,Leinweber:2002qb}. However, the
inclusion of the flavour singlet $\eta'$ does offer the opportunity
for detection of enhanced chiral behaviour in quenched simulations. As
a result of double-hairpin loops the leading nonanalytic contribution
is $\log\mpi$ --- i.e. magnetic moments diverge in the chiral limit
\cite{Savage:2001dy}.

In the case of the $\Delta$ it is the next-to-leading order
corrections which cause the most significant chiral effects. The
flavour symmetry of the $\Delta$ interpolating fields ensure that in
the quenched approximation, assuming isospin symmetry, the magnetic
moment is proportional to the charge. Just considering the
$\Delta^{++}$ is therefore sufficient to determine the chiral
behaviour of all charge states in the quenched theory.

We apply the diagrammatic method of Leinweber
\cite{Leinweber:2001jc,Leinweber:2002qb} to determine the quenched
chiral corrections to the $\Delta$ baryon magnetic moment.
Preliminary results have been documented in
Refs.~\cite{Leinweber:2003ux,Young:2003gd}.

As for the nucleon, the vertex correction from the double
hairpin $\eta'$ dressing gives rise to a leading logarithmic
divergence. The magnetic current induced by a pion loop is zero in the
quenched theory. In the case of the $\Delta^{++}$ this is a trivial
observation, as meson fields can only be generated by $u\bar{u}$-pairs
and must therefore be neutral.

It is the vertex corrections to the magnetic moment which give rise to
the most interesting physics in the quenched approximation. In
particular, it is once again the contribution from the intermediate
$uuu$-baryon which gives rise to a clear signal of quenched chiral
physics.  The contribution from the electromagnetic current coupling
to the baryon in Fig.~\ref{deltaFlow}(c) causes a reversal of the
chiral effect relative to physical QCD. As for the case of the
$\Delta$ mass as described above, the repulsive interaction of this
loop enhances the effect of quenching in the chiral regime.

\begin{figure}[t]
\includegraphics[width=\columnwidth]{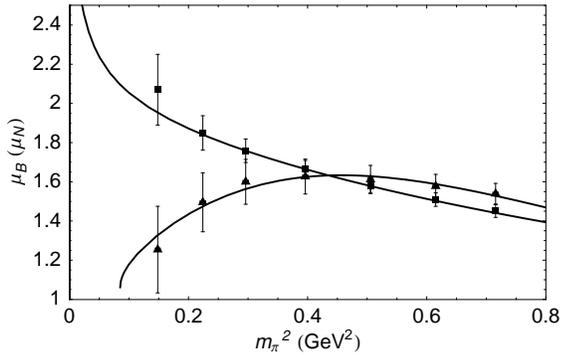}
\caption{Preliminary lattice results for proton ($\blacksquare$) and
  $\Delta^+$ ($\blacktriangle$) magnetic moments. Curves represent
  best fit to data using FRR Q\chiPT.}
\label{fig:mag}
\end{figure}

We show preliminary results of FLIC fermion simulations
\cite{Zanotti:2001yb} of baryon magnetic moments
\cite{Zanotti:2003gc}. In particular, we plot together the $p$ and
$\Delta^+$ magnetic moments in Fig.~\ref{fig:mag}.  We observe that
at moderate quark masses the expected heavy-quark theory result is
observed, with the $\Delta^+$ lying slighty above the proton.
However, in the light-quark regime chiral physics becomes apparent
with the clear separation of the two signals. The downward curvature
in the $\Delta^+$ is dominated by the coupling to an intermediate
octet baryon. This demonstrates a clear picture of nonperturbative
effects resulting from dynamical chiral symmetry breaking in quenched
lattice QCD.

Finite volume effects cannot play a significant role in the
conclusions drawn from this study.  The opposite sign on the
$\Delta\to N\pi$ loop means that such a vertex correction to the
$\Delta$ baryon magnetic moment will also have a sign which is
opposite to that of QCD.  The pion loop contributions to the magnetic
moments are $p$-wave and hence the main effect of the finite volume is
to create a gap in momentum space between $0$ and $2\pi/L$.  Because
the loop integrals are positive definite for both the proton and
$\Delta$ when $\mpi+m_N>m_\Delta$, the finite volume can only suppress
the magnitude of loop effects but cannot change their sign. Studying
the nucleon and $\Delta$ magnetic moments together therefore provides
a clear signal of quenched chiral physics independent of volume
effects.

Finite volume effects in the octet baryon sector have been studied in
more detail in Ref.~\cite{DBLCairns03}. It is found that the finite
volume discrepancies in the meson-loop contributions are negligible
at the pion masses of interest.

\section{CONCLUSIONS}
The study of both masses and magnetic moments in the quenched
approximation helps to provide a deeper understanding of chiral
extrapolation. This has also helped in developing a deeper
understanding of the internal structure of baryons. In particular, we
have identified that the chiral structures evaluated in FRR\chiPT\ are
able to describe quenched discrepancies in baryon mass calculations.

We identify that the $\Delta$ resonance offers unique features to
study the differing chiral properties of quenched and physical QCD.
The flattening of the $\Delta$ mass observed in FLIC fermion
simulations demonstrates a clear signal of quenching effects. In the
case of the magnetic moments the effect is much more dramatic. We see
the turnover in the $\Delta$ magnetic moment in the chiral regime
signifies clear nonperturbative effects, showing strong deviation from
the constituent-quark picture of QCD.

\section*{Acknowledgements}
This work was supported by the Australian Research Council.

\end{document}